\documentclass[preprint,12pt]{elsarticle}

\usepackage{amssymb}
\usepackage{amsmath}
\usepackage{graphicx}%
\usepackage{multirow}%
\usepackage{amsmath,amssymb,amsfonts}%
\usepackage{amsthm}%
\usepackage{mathrsfs}%
\usepackage[title]{appendix}%
\usepackage{xcolor}%
\usepackage{textcomp}%
\usepackage{manyfoot}%
\usepackage{booktabs}%
\usepackage{algorithm}%
\usepackage{algorithmicx}%
\usepackage{algpseudocode}%
\usepackage{listings}%

\usepackage{lineno}
\usepackage{xspace}
\usepackage{xcolor}
\usepackage{comment}
\usepackage{subfigure}
\usepackage{subcaption}

\newcommand{\jpsi} {\ensuremath{{\mathrm J}/\psi}\xspace}
\newcommand{\psip} {\ensuremath{\psi'}\xspace}
\newcommand{\rhozero} {\ensuremath{\rho^0}\xspace}

\newcommand{\starlight}{\textsc{STARlight}\xspace}

\newcommand{\nineH}        {$\sqrt{s}~=~0.9$~Te\kern-.1emV\xspace}
\newcommand{\seven}        {$\sqrt{s}~=~7$~Te\kern-.1emV\xspace}
\newcommand{\eight}        {$\sqrt{s}~=~8$~Te\kern-.1emV\xspace}
\newcommand{\twoH}         {$\sqrt{s}~=~0.2$~Te\kern-.1emV\xspace}
\newcommand{\twosevensix}  {$\sqrt{s}~=~2.76$~Te\kern-.1emV\xspace}
\newcommand{\five}         {$\sqrt{s}~=~5.02$~Te\kern-.1emV\xspace}
\newcommand{\fiveExactly}  {$\sqrt{s}~=~5$~Te\kern-.1emV\xspace}
\newcommand{\twosevensixnn}{$\sqrt{s_{\mathrm{NN}}}~=~2.76$~Te\kern-.1emV\xspace}
\newcommand{\fivenn}       {$\sqrt{s_{\mathrm{NN}}}~=~5.02$~Te\kern-.1emV\xspace}

\newcommand{\GeVc}         {Ge\kern-.1emV/$c$\xspace}
\newcommand{\MeVc}         {Me\kern-.1emV/$c$\xspace}
\newcommand{\TeV}          {Te\kern-.1emV\xspace}
\newcommand{\GeV}          {Ge\kern-.1emV\xspace}
\newcommand{\GeVtwo}       {Ge\kern-.1emV$^2$\xspace}
\newcommand{\MeV}          {Me\kern-.1emV\xspace}
\newcommand{\GeVmass}      {Ge\kern-.1emV/$c^2$\xspace}
\newcommand{\MeVmass}      {Me\kern-.1emV/$c^2$\xspace}

\journal{Computer Physics Communications}

\begin{document}

\begin{frontmatter}


\title{A fast Bayesian surrogate for the photon flux in ultra-peripheral collisions}

\author[1,2]{Simone Ragoni\corref{cor1}}
\ead{simone.ragoni@cern.ch}

\author[1]{Janet Seger}
\ead{jseger@creighton.edu}

\cortext[cor1]{Corresponding author}

\affiliation[1]{
  organization={Creighton University},
  addressline={2500 California Plz},
  city={Omaha},
  state={NE},
  postcode={68178},
  country={United States}
}

\affiliation[2]{
  organization={Institute of Physics, University of Silesia in Katowice},
  addressline={ul. Bankowa 12},
  city={Katowice},
  postcode={40-007},
  country={Poland}
}

\begin{abstract}
We present a fast surrogate for the Equivalent Photon Approximation (EPA) flux in ultraperipheral collisions (UPCs), based on a Bayesian neural network (BNN) trained over analytical flux integrals with an iterative procedure focused on regions of high relative uncertainties to minimise the number of integrations. The surrogate propagates  experimentally available uncertainties on the neutron skin thickness and surface diffuseness.  Once trained, this surrogate technique brings an estimated gain of two orders of magnitude in CPU time. The implementation provides a modular framework for rapidly propagating updated nuclear priors and assessing uncertainties for photon flux in future UPC analyses.
\end{abstract}

\begin{graphicalabstract}
\includegraphics[width=0.9\textwidth]{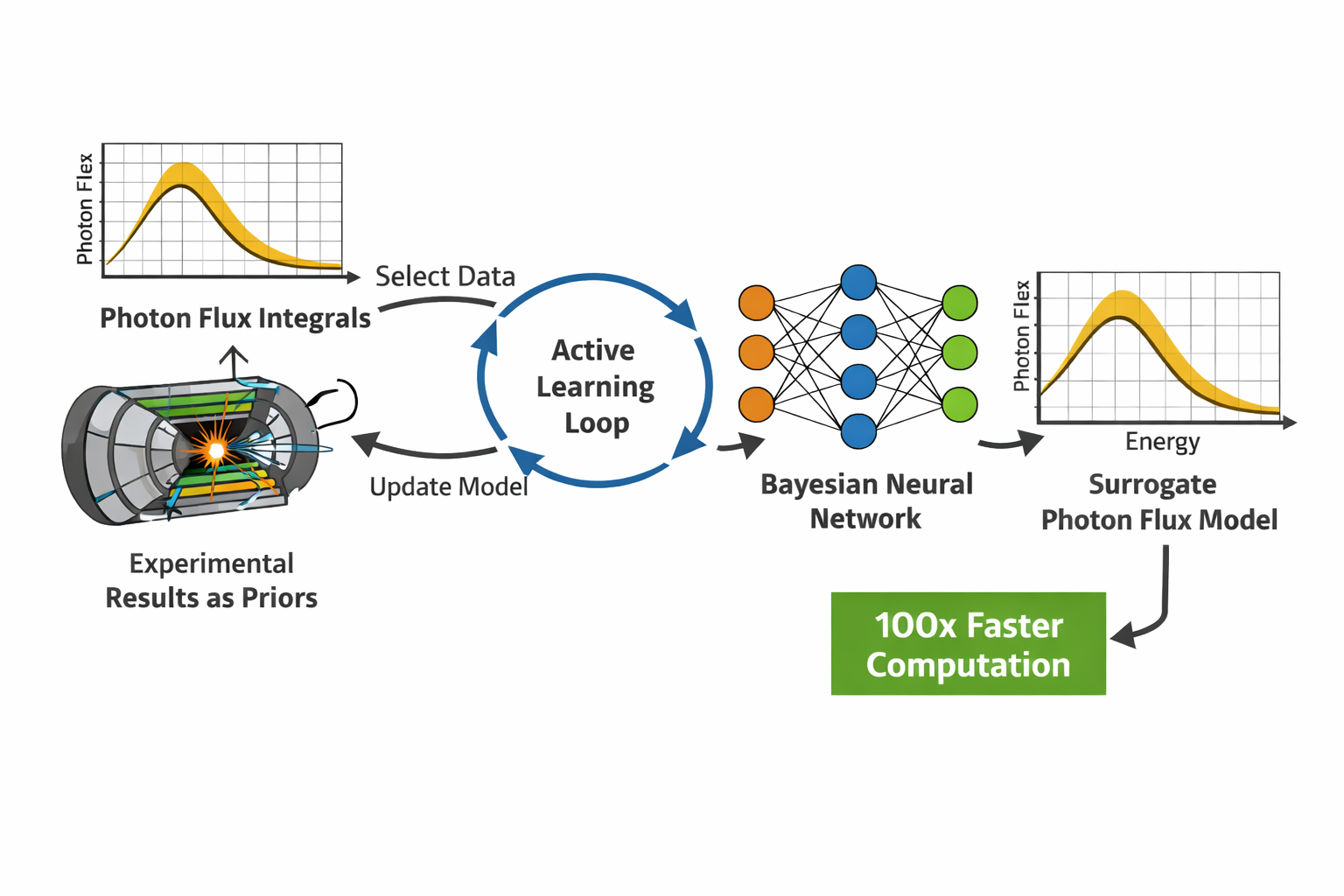}
\end{graphicalabstract}

\begin{highlights}
\item An active learning loop trains a Bayesian neural network surrogate.
\item Experimental constraints (priors) are incorporated during training.
\item Photon flux predictions are accelerated by up to two orders of magnitude.
\item The method naturally propagates uncertainties.
\end{highlights}


\end{frontmatter}

\section{Introduction}\label{sec:intro}
Ultraperipheral collisions (UPCs) at high‐energy colliders provide a unique environment to study photon‐induced processes in a clean setting, where the strong interaction is suppressed by requiring large impact parameters. In these interactions, the strong electromagnetic fields of the ultrarelativistic heavy ions act as \textit{quasireal} photon sources i.e. with very low virtualities, and a vector meson may be produced in the final state as a result of the interaction of the photon with the gluon distribution of the target. Vector meson photoproduction and exclusive lepton pair production have been studied by ALICE \cite{alice-vector-meson}, ATLAS \cite{ATLAS:2022ryk}, CMS \cite{cms-vector-meson}, and LHCb \cite{lhcb-vector-meson} at the CERN Large Hadron Collider (LHC), and STAR \cite{star-vector-meson} at the Relativistic Heavy Ion Collider (RHIC). In these measurements, the vector meson production cross sections are often related to the photonuclear cross sections by means of the photon flux model. This technique is needed to explore both nuclear shadowing and gluon saturation through \rhozero \cite{ALICE:2020ugp, STAR:2007elq}, \jpsi \cite{ALICE:2023jgu}, \psip \cite{Acharya:2748581, LHCb:2022ahs} production.  

The standard approach to photon‐flux calculation uses either the analytical Equivalent‐Density Form Factor (EDFF) formula employed in e.g. \starlight \cite{starlight} or a full two‐dimensional real‐space convolution of the point‐charge kernel with a Woods–Saxon nuclear density. Either of these involve computing several integrals at the launch of the program, to build look-up tables. While this cost is incurred only once per job, in typical high-energy-physics distributed computing environments thousands of independent jobs are executed, each repeating the same initialization procedure. As a consequence, identical photon-flux tables are recomputed many times, leading to a significant and unnecessary computational overhead.  

The solution we want to explore in this manuscript is to use a machine learning surrogate for the photon flux, which would mean that at run-time the photon flux is a simple call to a function from the trained surrogate, and thus very fast. We have chosen to realise it through  a physics‐informed Bayesian neural‐network (BNN). This technique has the advantage that the current experimental and theoretical knowledge regarding all aspects that lead to the calculation of the photon flux may be encoded quite naturally. Our surrogate then (i) incorporates priors on the neutron‐skin thickness  and Woods–Saxon diffuseness based on the latest experimental data \cite{PREX:2021umo, Klos:2007is}, (ii) employs Monte Carlo dropout to deliver uncertainty estimates, and (iii) uses an active‐learning strategy to focus expensive integrator calls on the most uncertain regions of parameter space. We release our code to provide a fast tool for future UPC analyses.

This approach has another benefit, since it provides a natural framework to estimate the uncertainty related to the modelling of the photon flux. This is crucial for measurements of the photonuclear cross section, where estimates of the flux uncertainty have typically been obtained by varying a single Woods–Saxon parameter (most often the nuclear radius) by an amount comparable to the nuclear skin (e.g. $\pm 0.5$ fm) and recomputing the flux \cite{ALICE:2018oyo}. While straightforward, this prescription neglects correlated uncertainties in other nuclear‐structure inputs (such as the diffuseness), which is where the model dependency of the photon flux originates.
Another possible strategy would be to perform scans in the parameter space to measure the uncertainty either through a similar variational technique or a Bayesian approach. This is however computationally expensive, since it would require to elaborate two‐dimensional integrals on a grid of multidimensional input vectors, either for systematic scans or for a traditional Bayesian propagation of the uncertainties. Our approach allows one to include many components to the uncertainty in the photon flux in a computationally efficient manner.

\section{Photon flux formalism}
\label{sec:epa}

In the Equivalent‐Photon Approximation (EPA), a fast‐moving nucleus is treated as a source of quasireal photons. The photon flux at impact parameter vector $\mathbf b$, and photon energy~$\omega$, is given by the two‐dimensional convolution of the point‐charge kernel with the projected nuclear density \cite{Baltz:2007kq}:
\begin{equation}
  N(\omega, b) \;=\;
  \int_{0}^{2\pi}\!d\phi
  \int_{0}^{R_{\rm nuc}} R\,dR\ 
  \rho_{2D}(R)\;
  n_{\rm pt}\bigl(\omega,\,|\mathbf C|\bigr)\,,
  \label{eq:epa_realspace}
\end{equation}
where $R_{\rm nuc}$ is a cutoff beyond which the nuclear density vanishes, $\mathbf R = (R\cos\phi,\;R\sin\phi)$, $\rho_{2D}$ is the projected two-dimensional density defined in Sec.~\ref{sec:ws}, and
\begin{equation}
    |\mathbf C| = |\mathbf b+\mathbf R| = \sqrt{b^2 + R^2 + 2\,b\,R\cos\phi} \text{ .}
\end{equation}

\subsection{Nuclear density model}
\label{sec:ws}
We adopt a Woods–Saxon form for the three‐dimensional nuclear density,
\begin{equation}
  \rho_{\rm WS}(r)
  = \frac{\rho_0}{1 + \exp\!\bigl((r - R_0)/a\bigr)}\,,
  \label{eq:ws3d}
\end{equation}
with radius parameter $R_0$ and surface diffuseness~$a$.  The normalization constant $\rho_0$ is fixed by
\[
  1 = \int d^3r\,\rho_{\rm WS}(r)
      = 4\pi \int_{0}^{\infty}\!dr\ r^2\,\frac{\rho_0}{1+\exp\!\bigl((r-R_0)/a\bigr)}\,.
\]
The projected two‐dimensional density entering Eq.~\eqref{eq:epa_realspace} is
\begin{equation}
  \rho_{2D}(R)
  = \int_{-\infty}^{+\infty} dz\;\rho_{\rm WS}\!\bigl(\sqrt{R^2 + z^2}\bigr)\,,
  \label{eq:proj2d}
\end{equation}
which we renormalize via
\[
  1 = \int_{0}^{\infty} 2\pi R\,dR\;\rho_{2D}(R)\,.
\]

\subsection{Point‐charge kernel}
The point‐charge EPA flux at fixed impact parameter $b$ is
\begin{equation}
  n_{\rm pt}(\omega,b)
  = \frac{Z^2\alpha}{\pi^2\,b^2}\,\xi^2
    \Bigl[K_1(\xi)^2 + \tfrac{1}{\gamma^2}\,K_0(\xi)^2\Bigr],
  \quad
  \xi \equiv \frac{\omega\,b}{\gamma}\,,
  \label{eq:point_flux}
\end{equation}
where $Z$ is the nuclear charge, $\alpha$ the fine‐structure constant, $\gamma$ the Lorentz factor, and $K_{0,1}$ the modified Bessel functions.

\section{Implementing a BNN surrogate}
\label{sec:bayesian-neural network}
The photon flux computed using Eq.~\ref{eq:epa_realspace} is used in this work to train the Bayesian neural-network. 
We implemented a Bayesian neural‐network (BNN) surrogate through \textit{Keras} \cite{chollet2015keras}, using the dropout technique highlighted in \cite{gal2016dropout,gal2016thesis}. The BNN takes as input the feature vector:
\begin{equation}
\mathbf x = \bigl[\ln\omega,\;b,\;\Delta R,\;a\bigr]\in\mathbb R^4,
\end{equation}
where $\ln\omega$ is the logarithm of the photon energy, $b$ the impact parameter, $\Delta R$ the shift of the nuclear radius relative to the Woods--Saxon baseline, and $a$ the nuclear surface diffuseness, with all inputs appropriately scaled for numerical stability during training.

The neural network is trained to predict the target
\begin{equation}
t \equiv \ln N(\omega,b),
\end{equation}
where $N(\omega,b)$ is the photon flux obtained from the numerical evaluation of Eq.~\eqref{eq:epa_realspace}.  

In this Bayesian implementation, the network does not return a single deterministic prediction, but rather an ensemble of stochastic outputs $t^{(k)}(\mathbf{x})$, obtained through Monte Carlo dropout \cite{gal2016dropout,gal2016thesis}. From this ensemble, we extract the predictive mean $\mu(\mathbf{x})$ and standard deviation $\sigma(\mathbf{x})$, which represent the surrogate prediction and its associated uncertainty, respectively.

The network parameters are optimized by minimizing the Gaussian negative log-likelihood loss applied to the target $t$, following the approach described in \cite{Kendall:2017tnb}.

In our BNN surrogate we model the prior for the shift of the nuclear radius as:
\begin{equation}
    \Delta R \sim \mathcal N(0.283,\,0.071^2)\,\mathrm{fm}\text{ ,}
    \label{eq:prior-r}
\end{equation}
where we take as mean and standard deviation the values measured by PREX-II in \cite{PREX:2021umo}, and $\mathcal N(\mu,\,\sigma^2)$ is the usual notation for a Gaussian with mean $\mu$ and sigma $\sigma$. We further use for the diffuseness the prior \cite{Klos:2007is}
\begin{equation}
    \Delta a \sim \mathcal N(0,\,0.015^2)\,\mathrm{fm}\text{ .}
    \label{eq:prior-a}
\end{equation}

\subsection{Active‐learning training loop}
\label{sec:active-learning}

A central part of this work is to minimise the number of expensive numerical integrations of Eq.~\ref{eq:epa_realspace} required to train the surrogate model, so as to be able to train it even on lightweight machines.   To achieve this, we employ an \emph{active‐learning} strategy \cite{gal2017deep}, in which the training set is iteratively refined to concentrate new samples in the regions of parameter space where the model exhibits the largest uncertainty.

At the start of the procedure, the Bayesian neural network (BNN) is trained on a small, randomly sampled \emph{seed} dataset $\mathcal D_0 = \{(\mathbf x_i, t_i)\}_{i=1}^{N_0}$ where $N_0$ is number of training points in that initial dataset. Subsequent rounds (\textit{active rounds}) follow an iterative cycle, which progressively refines the training set and drives convergence of the predictive uncertainty across the phase space of interest.

In each active round $r$, the following steps are executed:

\begin{enumerate}
  \item \textit{training step:} the BNN surrogate is trained for a fixed number of epochs (here, $10$) on the current dataset $\mathcal D_r$, minimizing the Gaussian negative log‐likelihood loss mentioned in  Sec.~\ref{sec:bayesian-neural network}. The resulting model provides both the mean $\mu(\mathbf x)$ and its uncertainty $\sigma(\mathbf x)$ for each point in the domain of the training set;

  \item \textit{candidate generation:} a set of $N_\mathrm{cand}$ new candidate points $\{\mathbf x_j'\}$ is generated across the kinematic domain. The candidates are drawn from a rapidity grid, where the conversion from photon energy to rapidity is via the well-known relation in DIS $y = \ln\left(2\omega/M_{VM}\right)$. The mass of the \jpsi $M_{\jpsi}$ is used in this context. The rapidity is then weighted by the uncertainty of the model from the previous round, such that
  \begin{equation}
      p(y) \propto \sigma_{r-1}(y) \text{,}
  \end{equation}
  favouring rapidity regions where the relative uncertainty is larger;

  \item \textit{uncertainty evaluation:} the trained model is applied to the new candidate set $\{\mathbf x_j'\}$, resulting in a new mean $\mu(\mathbf x_j')$ and uncertainty $\sigma(\mathbf x_j')$ for each point;

  \item \textit{sample selection:}
  the $k$ candidates with the highest predicted uncertainties are selected:
  \begin{equation}
          \mathcal S_r = \mathrm{arg\,top}_k \bigl(\sigma(\mathbf x_j')\bigr)\text{,}
  \end{equation}
  since these points represent the regions where the model has the largest uncertainties and are thus most likely to improve from additional evaluations (a larger training data set sampled more in that particular region);

  \item \textit{label acquisition (flux integration):} for each selected point $\mathbf x_j \in \mathcal S_r$, the target value $t_j = \ln N(\omega,b)$ is recomputed through the full  numerical integration of Eq.~\ref{eq:epa_realspace}.  These will complement the training data set from the previous round;

  \item \textit{dataset augmentation:} the new samples are added to the existing training data set:
  \begin{equation}
    \mathcal D_{r+1} = \mathcal D_r \cup
    \{(\mathbf x_j, t_j)\}_{j \in \mathcal S_r} \text{,}
  \end{equation}
  The model is then retrained on the expanded data set, and the process repeats;

  \item \textit{adaptive stopping:} the active‐learning loop continues until the maximum relative uncertainty predicted by the model across the rapidity grid satisfies a user‐defined threshold.  In our implementation, the loop stops automatically once the maximum relative uncertainty falls below $15\%$, ensuring a uniform precision level while avoiding unnecessary numerical integrations.
\end{enumerate}

This adaptive sampling procedure ensures that the computational budget is devoted primarily to those regions of the parameter space that are both physically relevant and difficult for the model to approximate. By focusing new training points where the epistemic uncertainty is largest, the active‐learning scheme allows the BNN surrogate to reach stable convergence with roughly two orders of magnitude fewer evaluations of the flux integral compared to uniform sampling.

\begin{figure*}[ht!]
  \begin{center}
    \subfigure[]{%
      \label{fig:al_r1_before}
      \includegraphics[width=0.47\textwidth]{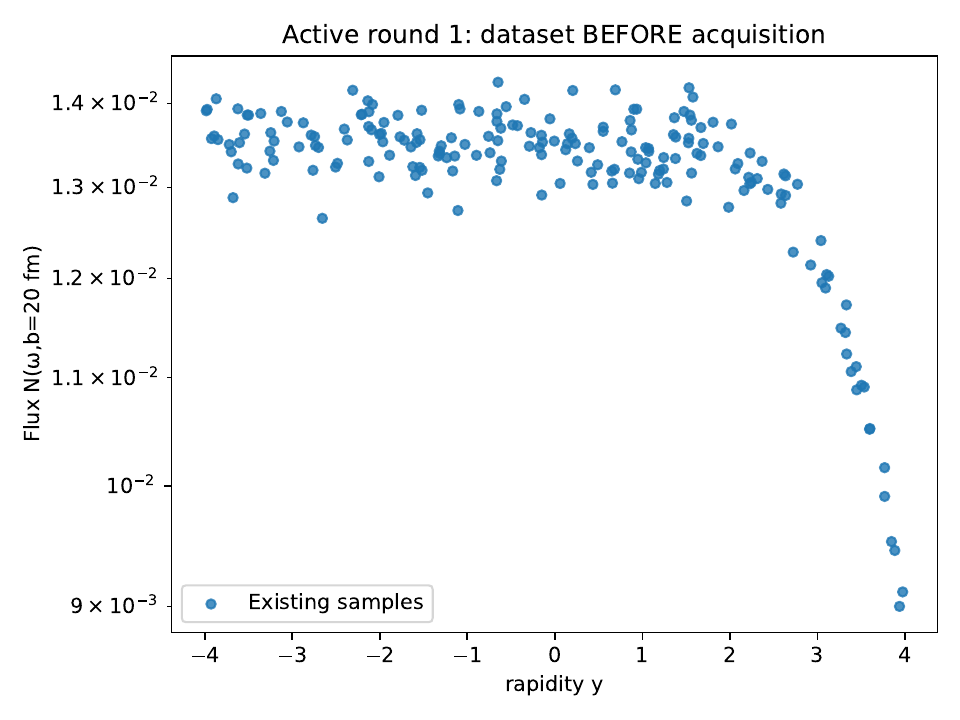}
    }
    \subfigure[]{%
      \label{fig:al_r1_unc}
      \includegraphics[width=0.47\textwidth]{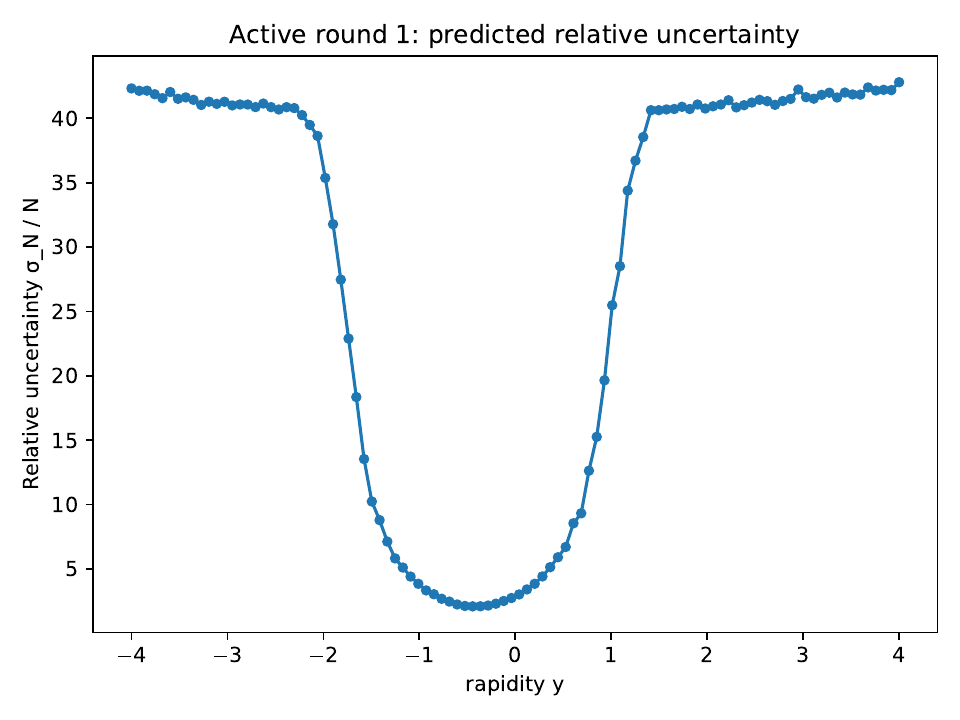}
    }\\ 

    \subfigure[]{%
      \label{fig:al_r2_before}
      \includegraphics[width=0.477\textwidth]{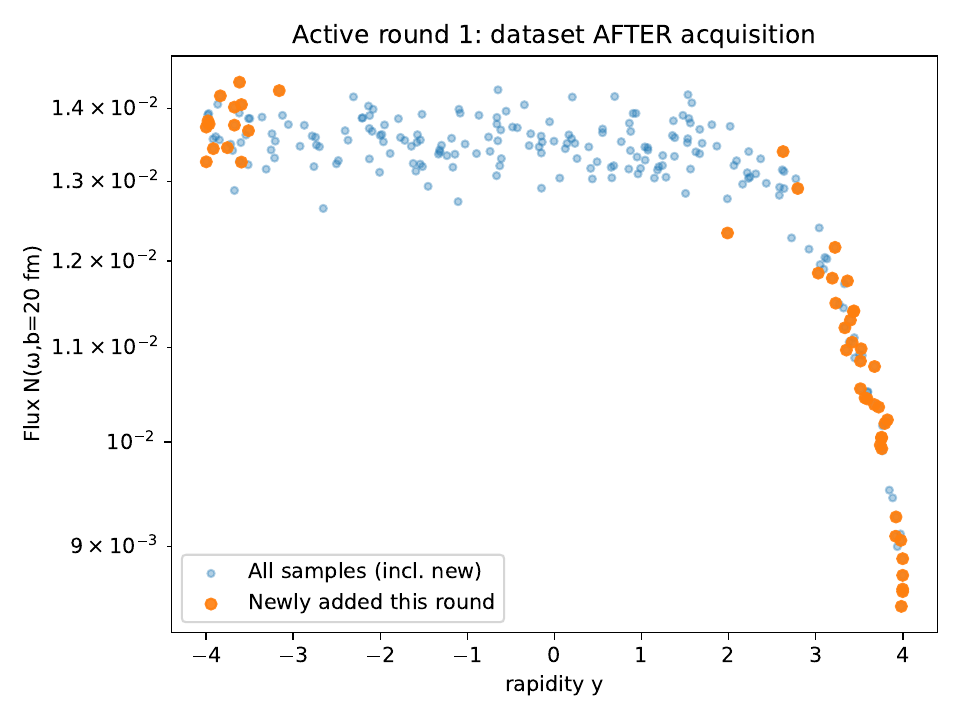}
    }
    \subfigure[]{%
      \label{fig:al_r2_unc}
      \includegraphics[width=0.47\textwidth]{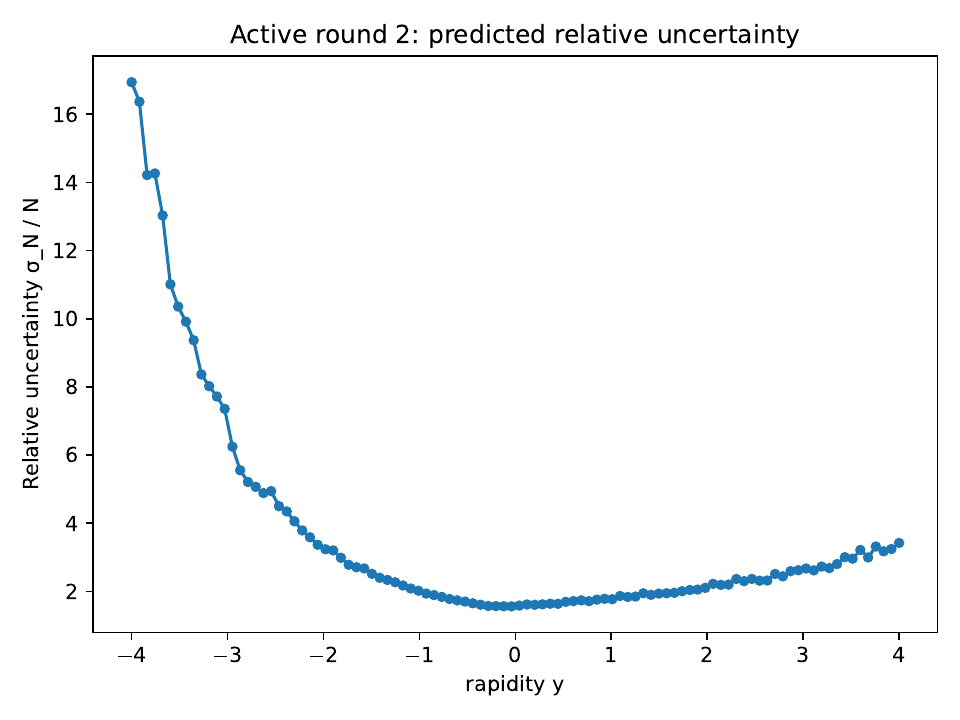}
    }\\ 

    \subfigure[]{%
      \label{fig:al_r3_before}
      \includegraphics[width=0.47\textwidth]{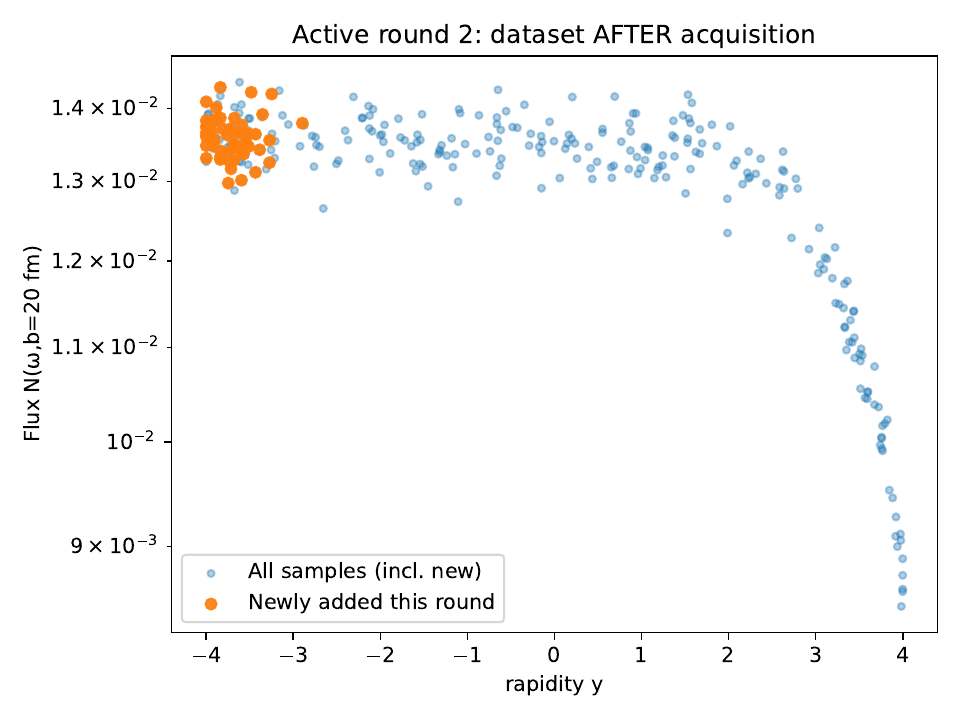}
    }
    \subfigure[]{%
      \label{fig:al_r3_unc}
      \includegraphics[width=0.47\textwidth]{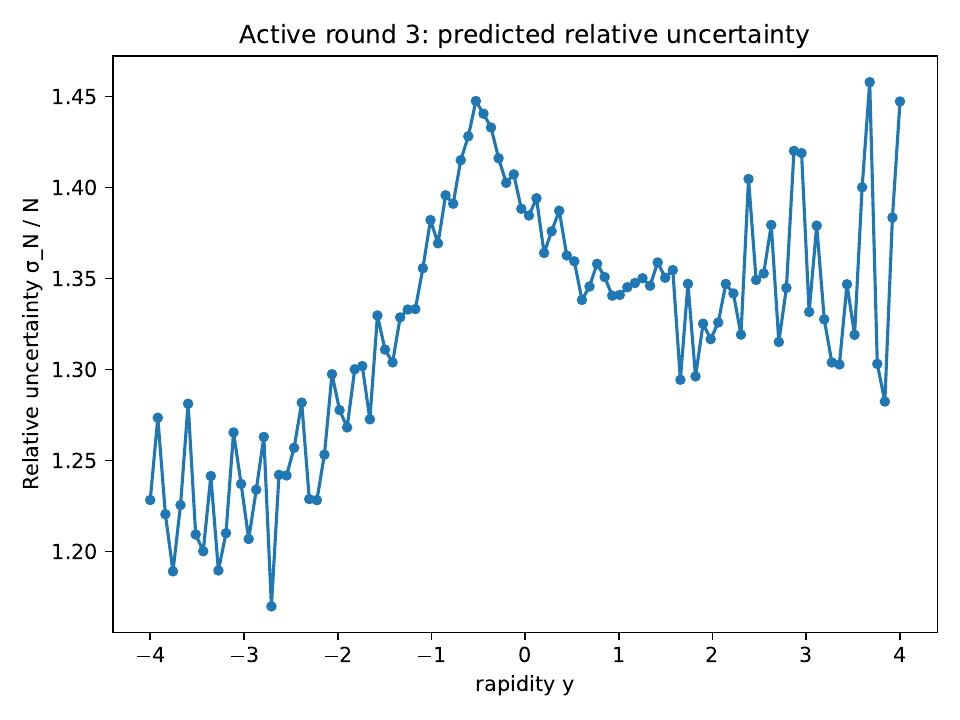}
    }\\ 

  \end{center}
  \caption{Figures~\ref{fig:al_r1_before}, \ref{fig:al_r2_before}, and \ref{fig:al_r3_before} show the training samples, before round 1, after round 1, and after round 2, respectively, used for the active-learning iterations, with the new incoming points in yellow, in regions of high uncertainty.   Figures~\ref{fig:al_r1_unc}, \ref{fig:al_r2_unc}, and \ref{fig:al_r3_unc} show the corresponding relative uncertainties after  each round. }
  \label{fig:al_seven_panel_subfigure}
\end{figure*}

\clearpage

\subsection{Convergence of the active learning loop}
\label{sec:convergence}

\begin{figure}[htbp]
  \centering
  \includegraphics[width=0.7\linewidth]{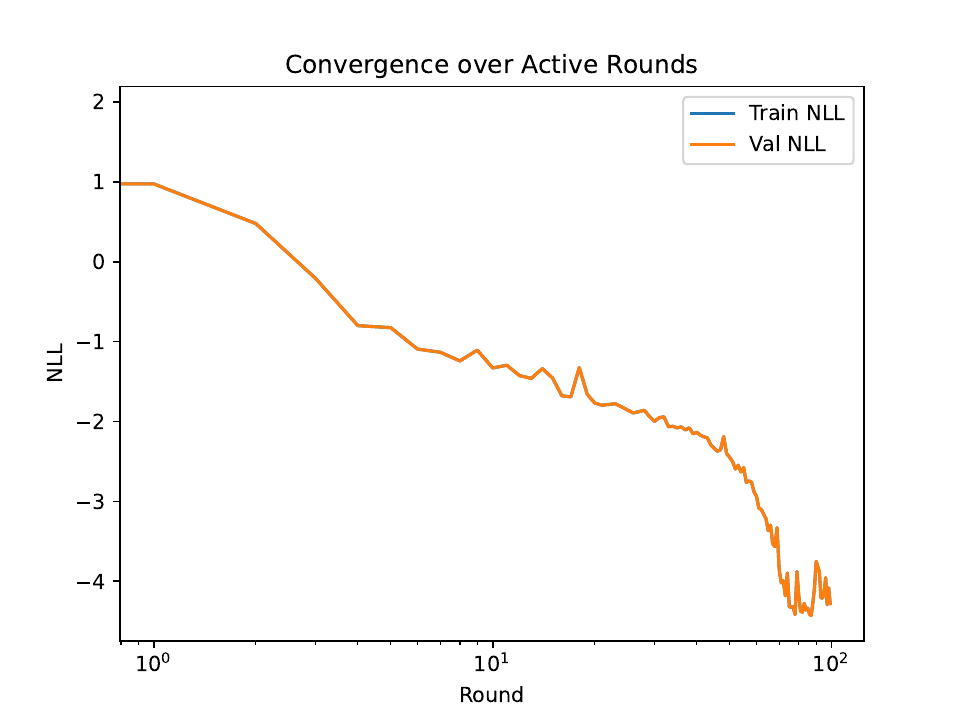}
  \caption{ Negative log‐likelihood (NLL) on the training set (blue) and on a held‐out evaluation set (orange) as a function of the active‐learning round. The horizontal axis is plotted on a logarithmic scale to emphasize early‐round behavior. }
  \label{fig:convergence}
\end{figure}
Figure~\ref{fig:convergence} demonstrates the robust, monotonic decrease in both the training and validation for the Gaussian negative log‐likelihood (NLL) as we iteratively augment our dataset via active sampling. Starting from positive values, it quickly becomes negative illustrating how the network is learning the overall structure of the photon flux mapping, before finally reaching convergence as it plateaus after round 70. The overall good overlap between training and validation curves shows there is minimal overfitting.

\section{Results}

\begin{figure}[htbp]
  \centering
  \includegraphics[width=0.7\linewidth]{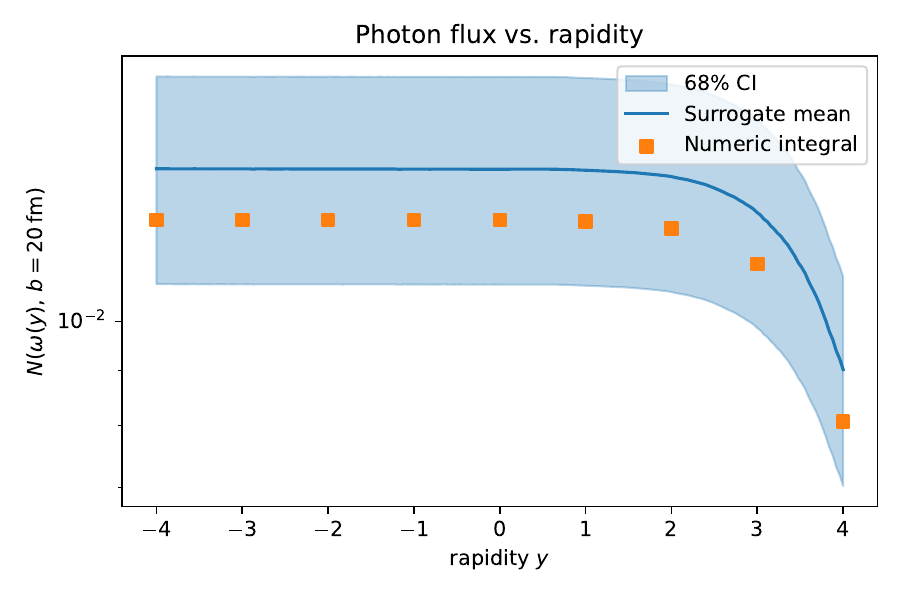}
  \caption{ Surrogate photon flux $N(\omega,b{=}20~\mathrm{fm})$ and $68\%$ credible interval compared to the direct numerical integration. }
  \label{fig:surrogate}
\end{figure}

Figure~\ref{fig:surrogate} shows the Bayesian‐NN surrogate prediction for the photon flux at fixed $b = 20$~fm as a function of rapidity, together with the points obtained from the direct numerical integrations of Eq.~\ref{eq:epa_realspace}. The shaded region represents the $68\%$ credible interval from the surrogate, while the solid line marks its median prediction. The agreement between the surrogate and the full numerical computation is excellent, with mean absolute residuals below $2\%$ across the entire rapidity range. This confirms that the active‐learning procedure efficiently allocates computational effort to regions of phase space that are more difficult to model, while maintaining precision elsewhere.

The model achieves a nearly uniform uncertainty band of $\sim 22$–$23\%$ between $y \approx -4$ and $y \approx +2$, with a mild rise at the highest rapidities due to the limited coverage of the training data at large photon energies. This $23\%$ uncertainty reflects the propagated PREX‐II priors on the neutron‐skin thickness and Woods–Saxon diffuseness. It is comparable in magnitude to the uncertainty quoted in \cite{ALICE:2018oyo}, where only the nuclear radius was varied, but arises here from a more complete Bayesian treatment that explicitly propagates the physical uncertainties on both radius and surface diffuseness.

The result demonstrates that, even when starting from the same physical priors, the Bayesian‐NN surrogate achieves comparable uncertainty estimates at a fraction of the computational cost. While the absolute uncertainty remains dominated by the poorly constrained neutron‐skin contribution, the method provides a framework to evaluate the uncertainty on the photon flux automatically whenever improved nuclear structure measurements become available.

The uncertainty on the photon flux propagates directly to the uncertainty on any photoproduction cross section extracted from  UPCs. If the flux uncertainty remains at the $\mathcal{O}(20\%)$ level, it will easily be the dominant systematic in all future  UPC studies. In this sense, the surrogate provides not only a computational speedup, but also a transparent and modular framework for systematically assessing and (potentially) reducing these uncertainties.

In fact, one of the main motivations for introducing this surrogate model is the substantial reduction in computational cost. In programs such as e.g. \starlight~\cite{starlight}, the integrals needed for the photon flux are recomputed each time a run is launched in order to build lookup tables for the production cross section as a function of rapidity and mass. As stated in the documentation of \cite{starlight}, this stage typically requires up to $\sim$15~minutes of single-core CPU time per configuration. In large-scale distributed computing campaigns, such as those required to produce large Monte Carlo samples, as typically done for experimental measurements by the LHC experiments, where thousands or tens of thousands of jobs are submitted per process with ${\cal O}(10^6$--$10^8)$ events, this initialization step is repeated on every node, corresponding to ${\cal O}(10^3$--$10^4)$~CPU~hours spent just for the integrations.

By contrast, the BNN surrogate requires a single instance of training (we have managed in $\sim$30~minutes on a single modern CPU core for ${\sim}2{\times}10^3$ training samples) and subsequently provides a fast predictions of $N(\omega,b)$ and its uncertainty, i.e. the time needed to run the input vector through the trained BNN. The time per evaluation is below the millisecond, and negligible compared to the event generation. This translates to a reduction in total CPU usage by three to four orders of magnitude for typical distributed Monte Carlo productions. As an additional benefit, since the surrogate delivers both central values and uncertainty estimates, it removes the need for repeated flux recalculations under parameter variations, which are needed for systematic studies.

\subsection{Customizing Nuclear Priors}
A key feature of the Bayesian‐NN surrogate is its modular treatment of the inputs from the nuclear structure studies. The model separates the physical priors on the Woods–Saxon parameters ($\Delta R$ for the thickness of the neutron skin and $a$ for diffuseness) from the neural network architecture itself. By default, these priors follow Eq.~\ref{eq:prior-r} and Eq.~\ref{eq:prior-a}, derived from the PREX‐II measurement of the neutron skin in $^{208}$Pb \cite{PREX:2021umo} and from the fits in \cite{Klos:2007is}. However, the framework allows users to replace these with custom priors or alternative nuclear density models.  

When new priors are provided, the active‐learning loop automatically regenerates the training dataset, retrains the surrogate, and propagates the updated uncertainties through to the final flux predictions. This modularity enables fast studies of how future experimental constraints on nuclear densities could reduce the systematic uncertainty originating from the photon flux, and hence their impact on future UPC measurements.

\section{Conclusion}
\label{sec:conclusion}

We have developed a fast surrogate for the EPA photon flux in ultraperipheral collisions based on a Bayesian neural network with MC‐dropout and active learning. The model provides results that compare within 2\% of the values obtained from numerical integrations as central values, while the 68\% band demonstrates a 23\% relative uncertainty at large rapidities. These uncertainties are obtained by propagating the current experimental uncertainties on parameters such the thickness of the neutron skin, measured by PREX‐II \cite{PREX:2021umo} and diffuseness \cite{Klos:2007is}, which are naturally incorporated as priors in the framework. These values are consistent with those obtained for current UPC measurements \cite{ALICE:2018oyo}.

The primary advantage of this method lies in its efficiency: once trained, the surrogate can evaluate $N(\omega,b)$ and its uncertainty nearly instantaneously, with roughly two orders of magnitude fewer calls to the full flux integrator. This makes it a practical tool for end‐to‐end uncertainty propagation in experimental analyses, as well as for rapid sensitivity studies exploring how improved nuclear‐structure measurements could tighten the systematic budget in photonuclear observables. If implemented for current models, such as \starlight \cite{starlight}, this would also enable much more computationally efficient simulations, relinquishing the need for expensive integrations on computing nodes at the level of distributed computing. 

Future extensions of this work will implement the treatment for correlated uncertainties across different nuclear observables, while also extending the currently available list of possible priors and source of uncertainties. This will allow for a unified description of systematics related to the photon flux, while also including the state-of-the-art for the currently available experimental and theoretical works.

\section*{Acknowledgements}

This work was funded by the Ministry of
Science and Higher Education of Poland (PL) and by the Department of Energy (DOE) of the United States of America (USA) through the grant DE-FG02-96ER40991.

\bibliographystyle{elsarticle-num} 
\bibliography{cas-refs}

\end{document}